\documentclass{emulateapj}
\shorttitle{Detectability of UCXBs as LISA sources}
\shortauthors{Chen, Liu \& Wang}

\begin{document}


\title{Detectability of ultra-compact X-ray binaries as LISA sources}


\author{Wen-Cong Chen$^{1,2}$, Dong-Dong Liu$^{3}$, and Bo Wang$^{3}$}
\affil{$^1$ School of Science, Qingdao University of Technology, Qingdao 266525, China;\\
 $^2$ School of Physics and Electrical Information, Shangqiu Normal University, Shangqiu 476000, China\\
 $^3$ Key Laboratory for the Structure and Evolution of Celestial Objects, Yunnan Observatories, Chinese Academy of Sciences, Kunming 650216, China;\\
 chenwc@pku.edu.cn, wangbo@ynao.ac.cn
}



\begin{abstract}
Ultra-compact X-ray binaries (UCXBs) are low-mass X-ray binaries with ultra-short orbital periods (usually less than 1 hour) and hydrogen-poor donor stars, which are proposed to be the potential LISA sources. In this work, we firstly employ the MESA code to examine the parameter space of the progenitors of UCXBs that LISA will detect. Our simulations indicate that the initial binaries with a neutron star and a $0.4-3.3~M_{\odot}$ companion star in an orbit of initial orbital period smaller than the bifurcation period could evolve into UCXBs, some of which will emit gravitational wave signals that can be detectable by the LISA. However, the initial orbital periods of the binaries that will evolve into UCXB-LISA sources in a distance of 10 kpc are located in a very narrow range, i.e. the formation of these LISA source requires an extremely fine-tuning of initial parameter. According to the characteristic stains and the derived maximum detectable distances, four sources among eight UCXBs with the observed distances are expected to be detected by the LISA. Based on the parameter space given by the detailed binary evolution models and the rapid binary star evolution code, the birthrate of UCXBs appearing as LISA sources in the Galaxy is estimated to be $(2-2.6)\times10^{\rm -6}~\rm yr^{-1}$. Considering the contribution of UCXBs in the globular clusters, the number of UCXB-LISA sources can reach $240 - 320$. Although the formation condition is severe, the detectability of UCXBs by the LISA is still optimistic and significant because they provide an opportunity to pursue full multi-messenger investigations.
\end{abstract}

\keywords{Gravitational wave sources (677); Gravitational waves (678); X-rays binary stars (1811);  compact binary stars (283); stellar evolution (1599); }

\section{Introduction}
The gravitational wave (GW) observation helped us to open a new window in understanding the Universe. It marked the start of astrophysical new era that the advanced LIGO detectors discovered the high-frequency GW signals from the double black hole (BH) merger event GW150914 in the distant galaxy \citep{abbo16}.  Another historic moment is the double neutron star (NS) coalescence event GW170817, which was detected throughout the entire electromagnetic spectrum from radio waves to gamma-rays, and became a milestone in opening multi-messenger astrophysics \citep{abbo17}.

Inspiral process of close binaries can emit low-frequency GW signals, which can provide some useful information in understanding compact binary evolution and binary interaction, including common envelope or envelope-ejection phase \citep{webb84,taam00,nele00,sluy06,ivan13}. A space GW interferometer Laser Interferometer Space Antenna (LISA) detecting low-frequency GW signals will launch in the early 2030s \citep{amar17}. The sensitive GW signals of LISA come from a frequency range between 0.1 mHz and 0.1 Hz, which can be produced by binary systems with orbital periods in the range of 20 s to 5 hours \citep{sluy11}. In the Galaxy, promising LISA sources include detached binaries such as double white dwarfs (WDs) \citep{hils00,nele01a,nele01c,nele03,yu10,kapl12,krem17,liu20}, WD - NS binaries \citep{taur18}, and double NSs \citep{yu15,taur17}, and interacting binaries like AM CVn stars \citep{nele01b,nele04b,nele03}, and ultra-compact X-ray binaries (UCXBs) \citep[][for reviews]{sluy11,nele13}.

UCXBs are a sub-population of low-mass X-ray binaries (LMXBs), which are characterized by ultra-short orbital
periods (usually less than 1 hour) and hydrogen-poor donor stars \citep{nels86,nele10} \footnote{Intermediate-mass X-ray binaries would firstly evolve into LMXBs, and subsequently bocome UCXBs \citep{pods02}.}. In the canonical model of UCXBs, a NS accretes material from a WD donor star in a binary with an orbital period less than one hour, in which the WD fills its Roche lobe due to the rapid orbital shrinkage induced by the gravitational radiation \citep{tutu79,tutu93,iben95,yung02,belc04,haaf12,seng17}. The pre-LMXBs consisting a NS and a main-sequence (MS) companion with initial orbital periods below the bifurcation period can also evolve into UCXBs \citep{pods02}.  Generally, the pre-intermediate-mass X-ray binaries (IMXBs) with initial orbital period much near the bifurcation period are easy to evolve toward ultra-compact orbit via the magnetic braking caused by the coupling between the magnetic field and an irradiation-driven wind \citep{chen16}. Other models such as NS + He star binaries \citep{hein13} or circumbinary disk around LMXBs/IMXBs \citep{ma09b} were also proposed as alternative evolutionary channels toward UCXBs. Due to the mass accretion, UCXBs provide a possibility to pursue full multi-messenger explorations in both GWs and electromagnetic waves. Recently, \cite{taur18} performed a systematic work on the detectability of NS + WD binaries by the LISA, and estimated that at least a hundred UCXBs will be detected by the LISA in the Galaxy. Interestingly, \cite{taur18} found that the WD masses in detached post-LMXBs consisting of a NS and a WD that can become visible LISA sources in the Hubble timescale are within an extremely narrow range ($0.162\pm0.005~M_{\odot}$), which provides an accurate constraint on the NS masses within $\sim 4\%$ errors by measuring the chirp signals from the detached pre-UCXBs.

In this work, we attempt to investigate the formation of UCXBs from LMXBs/IMXBs, and their detectability by the LISA in the Galaxy. We firstly explore the parameter space of the progenitor of UCXBs using the detailed binary evolution models in Section 2. The formation and detectability of eight UCXBs with the observed distances are explored in Section 3. Using the rapid binary star evolution (BSE) code, the number and birthrate of UCXBs appearing as potential LISA sources in the Galaxy are calculated in Section 4. Finally, we present a brief discussion and conclusion in Section 5.

\section{Formation and Evolution of UCXBs}
\subsection{Evolutionary code}
We employ a MESAbinary update version (r12115) in the Modules for Experiments in Stellar
Astrophysics code \citep[MESA;][]{paxt15} to simulate the evolution of LMXBs/IMXBs and the formation of UCXBs.
All UCXBs are assumed to evolve from the pre-LMXBs/IMXBs consisting of a NS (with a mass of
$M_{\rm ns}$) and a MS companion star (with a mass of $M_{\rm d}$).  The orbits of the binaries
are assumed to be circular and synchronized. The NS is thought to be a point mass, and the chemical composition of the companion star is similar to solar composition (X= 0.7, Y= 0.28, Z = 0.02). The input parameter space for the initial companion-star masses $M_{\rm d,i}$ and the initial orbital periods $P_{\rm orb,i}$ are set to be $0.4-3.5~M_{\odot}$ and $0.1 - 4$ days, respectively. In this work, the X-ray binaries are named to be LMXBs, and IMXBs when the initial donor-star masses are in the range of $0.4- 2.0~M_{\odot}$, and $2.1- 3.5~M_{\odot}$, respectively. Actually, IMXBs appear as LMXBs in most of their X-ray active lifetime \citep{pods02}. Even if the donor-star masses of IMXBs decrease to be less than $2.0~M_{\odot}$ during the mass transfer, they are still named as IMXBs for simplicity.

The orbital evolution of the binary system is dominated by three loss mechanisms of angular momentum including gravitational radiation, magnetic braking \citep{rapp83}, and mass loss. If the donor star possesses both a convective envelope
and a radiative core, magnetic braking is arranged to operate (magnetic braking index $\gamma=4$).
Once the donor star fills its Roche lobe, it will transfer the material onto the NS at a rate $\dot{M}_{\rm tr}$.
For the mass transfer efficiency, we adopt a model given by \cite{taur06} with $\alpha=0,\beta=0.5,\delta=0$, where $\alpha$
is the fraction of mass lost from the vicinity of the donor star, $\beta$ is the fraction of mass lost from the vicinity of the NS, and $\delta$ is the fraction of mass lost from circumbinary co-planar toroid.
If the accretion rate onto the NS $\dot{M}_{\rm ac}=(1-\beta)\dot{M}_{\rm tr}$ exceeds the Eddington accretion rate \citep{pods03}, the excess mass transfer rate is assumed to be ejected from the vicinity of the NS, carrying away the specific
orbital angular momentum of the NS. For simplicity, the irradiation effect of X-ray luminosities of accreting NS is not included \citep[see also][]{lu17}. We run the MESA code until the stellar age is greater than the Hubble timescale.

Once the LMXB/IMXB evolves into an UCXB with orbital period $P_{\rm orb}\la1-2$ hours, it will have a possibility to be detected by the LISA. Considering an observation time $T=4~\rm yr$, the characteristic strain of UCXBs can be expressed as \citep{chen20}
\begin{eqnarray}
h_{\rm c}\approx 3.75\times 10^{-20}\left(\frac{f_{\rm gw}}{1~\rm mHz}\right)^{7/6}\left(\frac{\mathcal{M}}{1~M_{\odot}}\right)^{5/3}\nonumber\\
\left(\frac{10~\rm kpc}{d}\right),
\end{eqnarray}
where $f_{\rm gw}=2/P_{\rm orb}$ is the GW frequency, $d$ is the distance of the source. The chirp mass is \citep{taur18}
\begin{equation}
\mathcal{M}=\frac{c^{3}}{G}\left(\frac{5\pi^{-8/3}}{96}f_{\rm gw}^{-11/3}\dot{f}_{\rm gw}\right)^{3/5},
\end{equation}
where $G$ is the gravitational constant, $c$ is the speed of light in vacuo, $\dot{f}_{\rm gw}$ is the GW frequency derivative.
Some previous studies adopted different critical orbital periods of UCXBs such as 60 \citep{ma09b,haaf12,haaf13,seng17}, and 80 minutes \citep{pods02,cart13,hein13}. Here, we extend the critical period of UCXBs to be 90 minutes. If the calculated characteristic strain is greater than the LISA sensitivity, the corresponding UCXBs are thought to be LISA sources.

\begin{figure}
\centering
\includegraphics[width=1.15\linewidth,trim={0 0 0 0},clip]{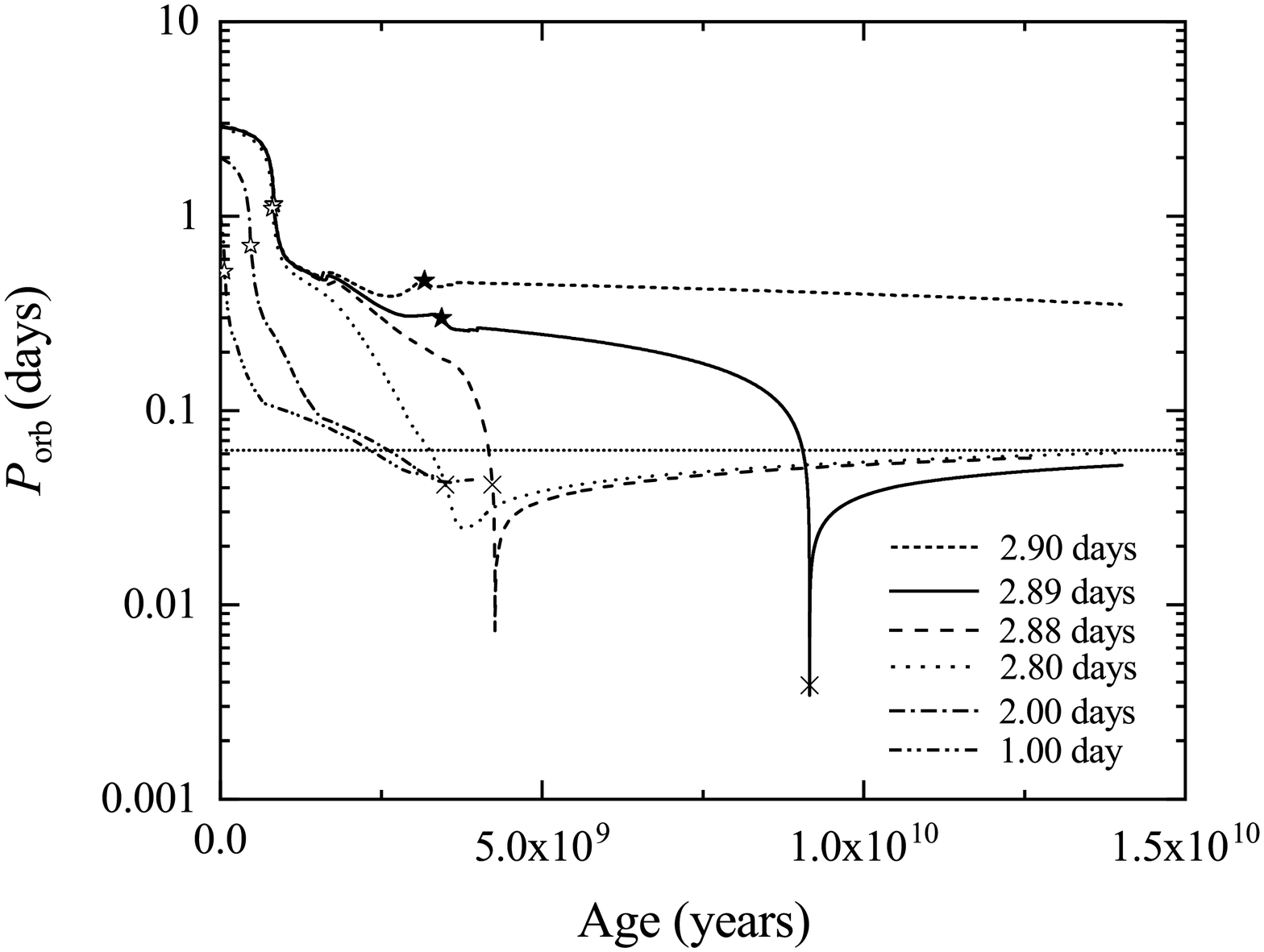}
\caption{Evolutionary tracks of pre-LMXBs with a donor star mass of $2.0~\rm M_{\odot}$ and different initial orbital periods in the orbital period vs. stellar age diagram. The open stars, solid stars, and crosses denote the onset of LMXBs, the end of LMXBs, and the onset of UCXBs, respectively. The horizontal short-dotted line represents an orbital period of 1.5 hours.} \label{fig:orbmass}
\end{figure}

\subsection{Simulated results}
To explore the formation of UCXBs, we calculate the evolution of pre-LMXBs consisting of a $1.4~\rm M_{\odot}$ NS and a $2.0~\rm M_{\odot}$ MS companion star for an initial orbital period of 1.0 - 3.0 days. In Figure 1, we plot the evolutionary tracks of six pre-LMXBs in the orbital period versus stellar age diagram. Due to relatively close orbit, magnetic braking efficiently causes the orbital separation to shrink, and the companion star can fill its Roche lobe within 0.5 - 1.0 Gyr for an initial orbital period of 2.0 -3.0 days. However, the evolutionary timescale before Roche lobe overflow is only 65 Myr when the initial orbital period is 1.0 day. The secular evolution of LMXBs/IMXBs depends on their initial orbital periods. The LMXBs/IMXBs would evolve toward compact binaries if the initial orbital periods are less than a critical period (otherwise would produce long orbital-period systems), which is the so-called bifurcation period \citep{tutu85}. \cite{sluy05a,sluy05b} explored the bifurcation period of LMXBs evolving toward UCXBs, and defined the bifurcation period as the longest initial period of pre-LMXBs evolving into UCXBs within
the Hubble timescale. The bifurcation period is very sensitive to the angular-momentum-loss mechanisms including the magnetic braking law \citep{pyly88,ma09a}, and the mass-loss mechanisms \citep{ergm96a,ergm96b,pods02}. Based on the angular momentum loss mechanisms adopted in this work, the bifurcation period of the pre-LMXB with a donor star of $M_{\rm d}=2.0~M_{\odot}$ is 2.89 days. Our simulated results are consistent with the conclusion given by \cite{chen16}, the pre-LMXBs with initial periods slightly smaller than the bifurcation period tend to obtain a relatively small minimum orbital period. For example, when the initial period is $P_{\rm orb,i}=2.89$ days, the LMXB will evolve into an UCXB with minimum orbital period $P_{\rm orb,min}=4.9$ min. The reason causing this phenomenon is as follows: a long initial orbital period would require a long evolutionary timescale, resulting in a high He abundances in the donor-star core, which subsequently produces more compact donor star and correspondingly shorter orbital period \citep{tutu87,lin11}. During the UCXB stage, the angular momentum loss is dominated by the gravitational radiation, which is at least 1-2 orders of magnitude higher than the magnetic braking. For an initial orbital period larger than 2.9 day, the LMXB will evolve into a binary millisecond pulsar with a He WD companion and a wide orbit \citep[e.g.][]{taur00,shao12}. The pre-LMXBs with initial orbital period of 2 - 2.89 days are the promising progenitors of UCXBs with orbital periods $P_{\rm orb}< 1~\rm hours$, while the minimum orbital period only reach 1.1 hour for pre-LMXB with an initial orbital period of 1.0 days. The main reason is that the H abundance in the center of the donor stars exceeds 0.45, which are difficult to reach a relatively compact state.

Figure 2 shows the evolutionary tracks of pre-LMXBs in the mass transfer rate versus the stellar age. The left panel illustrates the evolutionary examples of pre-LMXBs consisting of a $M_{\rm ns}=1.4~M_{\odot}$ NS and a $M_{\rm d}=2.0~M_{\odot}$ companion star. After the nuclear evolution of 0.83 Gyr, the donor star fills its Roche lobe, and transfers the surface H-rich material onto the NS. For an initial orbital period of 2.89 days, the pre-LMXB will experience three stages including LMXB, post-LMXB (or pre-UCXB), and UCXB stages. When the stellar age $t=3.45$ Gyr, the LMXB evolves into a detached post-LMXB consisting a radio millisecond pulsar (MSP) and a He core with a mass of $0.17~M_{\odot}$ in an orbit of 7.1 hours. During $t=3.45-9.16~\rm Gyr$, the post-LMXB appears as a binary millisecond radio pulsar. In this stage, the low-mass He core firstly evolves into a He WD after a $\sim 2~\rm Gyr$ contraction phase, and then begins a cooling track \citep{istr14a}. Because the system gradually spirals due to the angular momentum loss driven by GW radiation, it emits low-frequency GW signals in the final stage of cooling phase \citep{taur18}. Subsequently, the WD fills the Roche lobe, and triggers the second mass transfer when the orbital period is 0.1 hour. In this stage, the NS appears as a X-ray MSP, and the binary can be observable as an UCXB. It is very sensitive with the initial orbital period whether the pre-LMXBs/IMXBs can evolve into a detached pre-UCXB including a MSP and a He WD. When the initial orbital period is 2.88 days, the system can still form UCXB, while it always experiences mass transfer (see also the dashed curve in the left panel of Figure 2) without a detached stage. It seems that the pre-LMXBs/IMXBs require a fine-tuning initial orbital period (approximately equal to the bifurcation period) to form detached pre-UCXBs \citep[see also][]{istr14b}. It is worth noting that the pre-LMXBs/IMXBs would directly evolve into UCXBs without experiencing a detached pre-UCXBs stage if the companion masses are $\leq 1.5~M_{\odot}$ (see also the right panel of Figure 2) or the initial orbital periods are less than the bifurcation periods.

\begin{figure*}
\centering
\centering
\begin{tabular}{cc}
\includegraphics[width=0.48\textwidth,trim={30 10 30 30},clip]{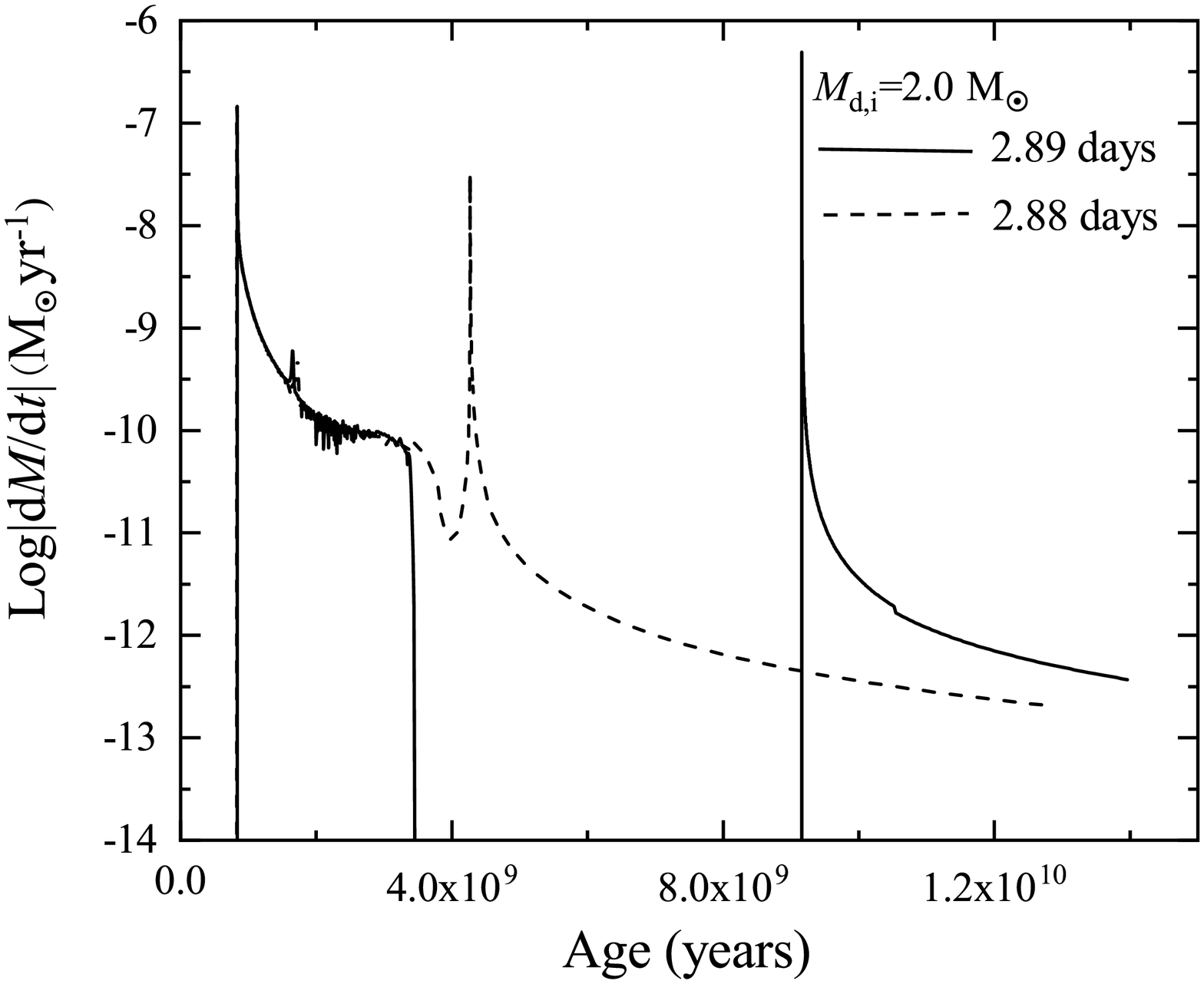} &    \includegraphics[width=0.48\textwidth,trim={30 10 30 30},clip]{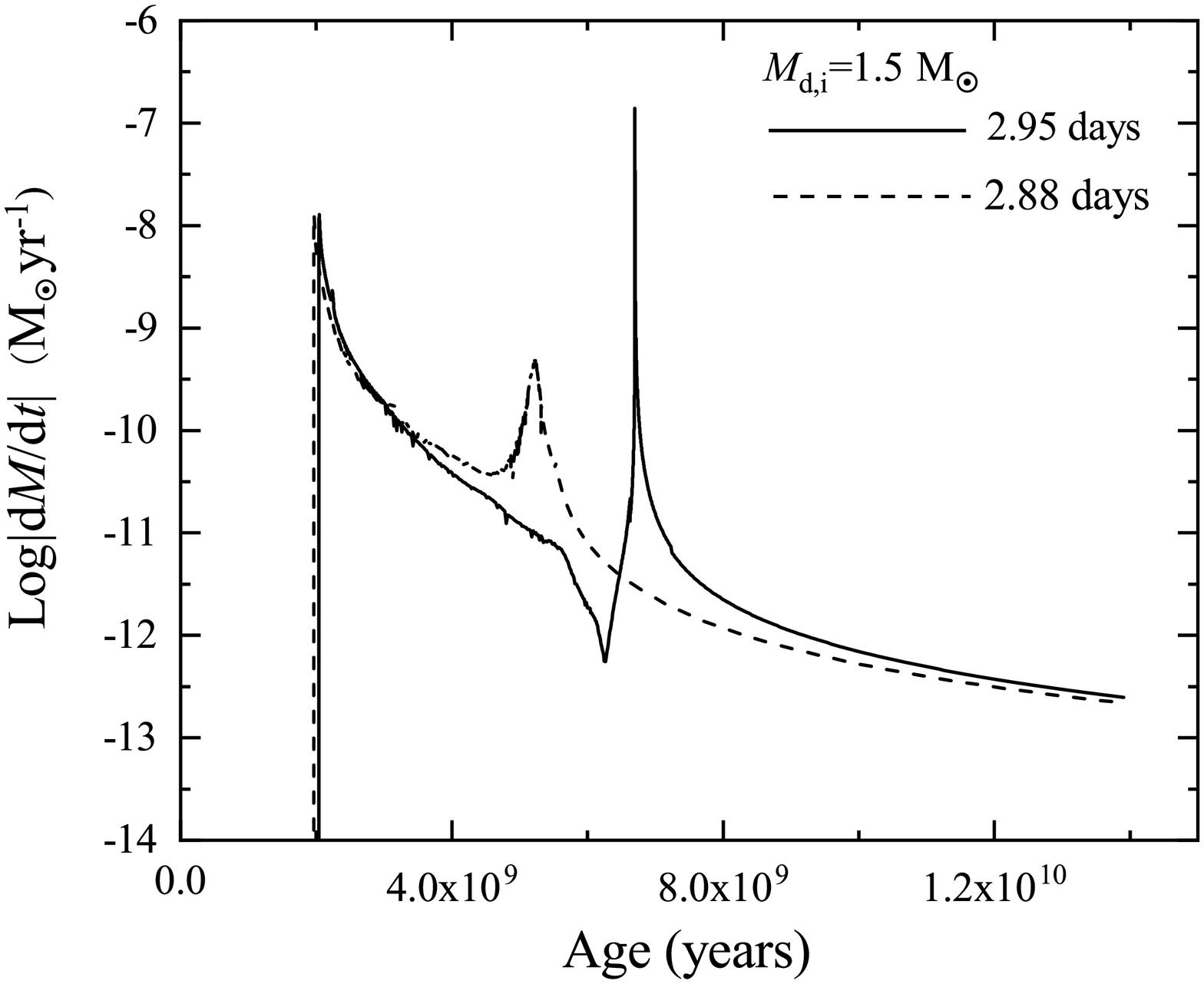} \\
\end{tabular}
\caption{Evolution of the mass transfer rate of the donor star for LMXB/IMXB in the mass transfer rate vs. stellar age diagram. The initial masses of the companion stars in the left and right panels are 2.0, and $1.5~M_{\odot}$, respectively.} \label{fig:orbmass}
\end{figure*}

Table 1 lists some main evolutionary parameters of pre-LMXBs/IMXBs that can evolve into detached pre-UCXBs consisting of a radio MSP and a He WD. The He WD masses in pre-UCXBs emerge within a narrow range of $0.160-0.170~M_{\odot}$, which is similar to the results given by \cite{taur18}. It is very sensitive with the initial NS mass, the accretion efficiency, and magnetic braking index $\gamma$ whether the pre-LMXBs/IMXBs would evolve toward a detached pre-UCXB. When the accretion efficiency is 0.3 (our accretion efficiency is 0.5), a pre-LMXB with $1.3~M_{\odot}$ NS orbiting a $1.4~M_{\odot}$ MS star in an orbit of 3.0 days would experience a detached stage \citep{taur18}. For a high magnetic braking index $\gamma=5$ \citep{istr14b}, pre-LMXBs with similar masses and a relatively wide range of initial orbital periods can evolve into detached pre-UCXBs \citep[see also Figure 3 in][]{seng17}. For UCXBs that originate from detached pre-UCXBs consisting of a NS + He WD, the minimum orbital periods are in the range of 5 - 6 min, which are similar to that of the simulated results given by \cite{seng17}.

\begin{figure*}
\centering
\includegraphics[width=2.0\columnwidth]{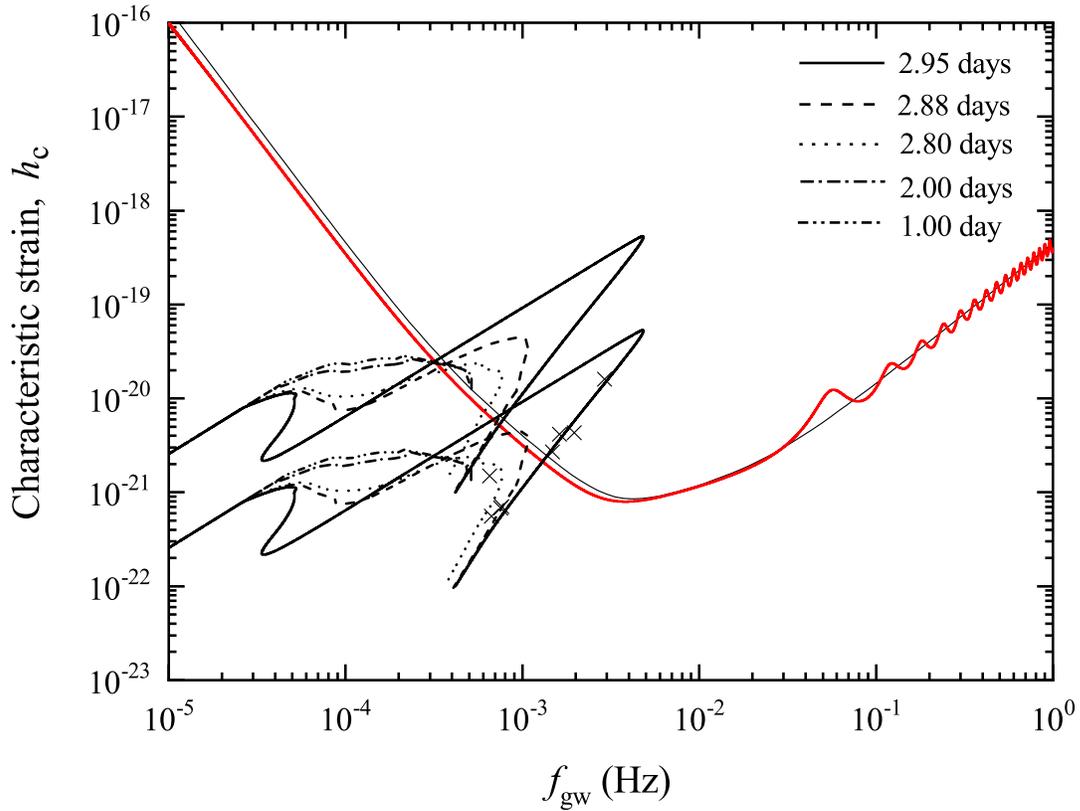}
\caption{Evolutionary tracks of five pre-LMXBs with a $1.4~M_{\odot}$ NS and a $1.5~M_{\odot}$ MS companion in the characteristic strain amplitude vs. GW frequency diagram. The red curve is the LISA sensitivity curve based on 4 yr of observations. The solid, dashed, dotted, dashed-dotted, and dashed-dotted-dotted curves correspond to the evolutionary tracks of pre-LMXBs with initial orbital periods $P_{\rm orb}=2.95,2.88,2.8,2.0$, and $1.0~\rm days$, respectively. The upper and under curve groups represent the distance of 1 kpc and 10 kpc, respectively. The crosses denote eight UCXBs with the observed distances.} \label{fig:orbmass}
\end{figure*}

Figure 3 illustrates the evolutionary tracks of pre-LMXBs consisting of a NS  and a $1.5~M_{\odot}$ donor star with five different initial orbital periods in the characteristic strain amplitude versus GW frequency diagram. The black curve represents the LISA sensitivity curve originating from a good analytic estimation (see also equation 13 in Robson et al. 2018), while the red curve arises from the numerical calculation. When $d=1$ kpc, five pre-LMXBs with an initial orbital period smaller than the bifurcation period can penetrate the LISA sensitivity curve, and are detectable by the LISA. However, only pre-LMXBs with orbital periods in a narrow range of $2.88-2.95~\rm days$ can be visible as the LISA sources within a distance of 10 kpc.

To explore the initial parameter space of the progenitors (pre-LMXBs/IMXBs) of UCXB-LISA sources, we have modeled the evolution of a great number of LMXBs/IMXBs. Figure 4 summarizes the final fates of LMXBs/IMXBs in the $P_{\rm orb,i}-M_{\rm d,i}$ plane, the solid stars denote the progenitors of UCXBs that will be visible as LISA sources within a distance $d=1$ kpc. The solid curve shows the bifurcation periods of pre-LMXBs/IMXBs with different initial donor-star masses. All systems above this curve will evolve into binary systems with long orbital periods. The pre-LMXBs/IMXBs with initial orbital periods much near the bifurcation periods tend to form UCXBs with extremely short orbital periods, and appear as visible LISA source within a long distance. The dashed curve represents the minimum initial orbital periods of the progenitors of UCXB-LISA sources in a distance of 10 kpc, i.e. the pre-LMXBs/IMXBs with initial parameters in the region between these two curves would evolve into UCXBs that can be detectable by the LISA at 10 kpc. None of UCXBs originating from pre-LMXBs with donor-star masses $M_{\rm d}\leq 1.0~M_{\odot}$ can be detectable by the LISA in a distance of 10 kpc. It is clear that the progenitors of these LISA sources require an extremely fine-tuning of initial orbital periods. The pre-LMXBs with initial orbital periods obviously lower than the bifurcation periods will also evolve into UCXBs, while they are only visible by the LISA in a close distance ($\sim1~\rm kpc$).

\begin{figure*}
\centering
\includegraphics[width=2.0\columnwidth]{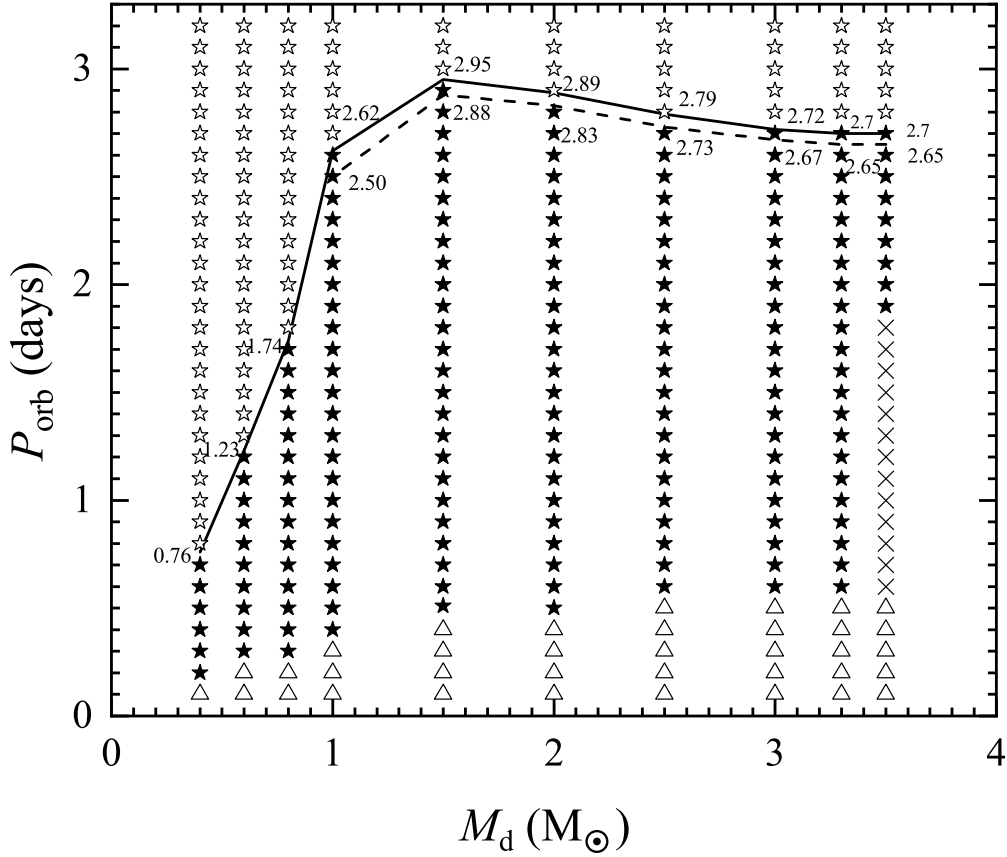}
\caption{Parameter space distribution of the pre-LMXBs/IMXBs evolving into UCXBs
in the initial orbital period vs. initial donor-star mass diagram.
The initial masses of NS are assumed to be $1.4~M_{\odot}$. The solid curve represents the bifurcation periods of LMXBs/IMXBs with different donor star masses, while the dashed curve denote the minimum initial orbital period of the progenitors of UCXBs that can be detected by the LISA within a distance $d=10~\rm kpc$. The solid stars represent the pre-LMXBs that can evolve into UCXBs, which are visible by the LISA within a distance $d=1~\rm kpc$. The open stars represent the pre-LMXBs/IMXBs evolving toward systems with long orbital periods. The open triangles denote the binary systems with donor stars that have already filled their Roche lobe at the beginning of binary evolution, and the crosses denote the binary systems experiencing a unstable mass transfer. Numbers inside the solid and dashed curves denote the initial orbital periods in units of days.} \label{fig:orbmass}
\end{figure*}

\begin{table*}
\begin{center}
\caption{Selected Evolutionary Properties for UCXBs and their progenitors for different initial donor star masses and initial orbital periods. \label{tbl-2}}
\begin{tabular}{@{}lllllllllll@{}}
\hline\hline\noalign{\smallskip}
$M_{\rm d,i}$ & $P_{\rm i, orb}$ &  $t_{\rm rlov}$ &$t_{\rm deta}$ &$P_{\rm deta}$ &$M_{\rm wd}$&$t_{\rm ucxb}$ &$P_{\rm ucxb}$&$P_{\rm min}$ & $f_{\rm i,LISA}$ &$\bigtriangleup t_{\rm LISA}$ \\
 ($ M_{\odot}$)     &  (days)  & (Gyr)   & (Gyr)  & (days)    &  ($ M_{\odot}$ )  &  (Gyr)  & (days)& (min) & (mHz)& (Myr)\\
\hline\noalign{\smallskip}

2.0 & 2.89 &0.83&3.45& 0.297&0.170&9.16 &0.004 & 4.94 & 0.83&34.9\\

2.5 & 2.79 &0.44&2.89&0.319 &0.166&9.49 &0.004 & 5.12 &0.84 &34.8\\

3.0 & 2.72 &0.26&2.69&0.274&0.160& 6.02& 0.005 & 5.83 &0.87 &14.0\\

3.3 & 2.70 &0.20&2.43&0.347&0.165&10.86 &0.004  & 5.14 &0.87&34.6\\
\hline\noalign{\smallskip}
\end{tabular}
\tablenotetext{}{}\\{Note. The columns list (in order): the initial donor-star mass, the initial orbital period, the stellar age at the beginning of Roche lobe overflow, the stellar age and the orbital period when the binary becomes a detached system, the WD mass, the stellar age and the orbital period when the system appears as a UCXB, the minimum orbital period, the initial GW frequency that the binary is detectable by the LISA, and the timescale that the binary appears as LISA source.}
\end{center}
\end{table*}
\section{Observations}
\subsection{The WD mass-orbital period relation}

\begin{figure}
\centering
\includegraphics[width=1.15\linewidth,trim={0 0 0 0},clip]{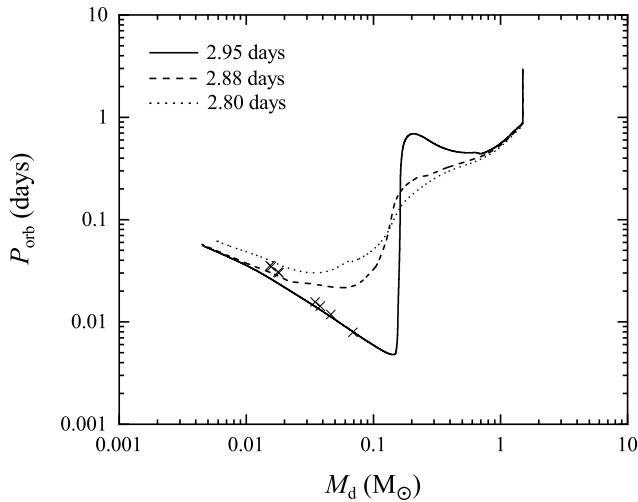}
\caption{Evolutionary tracks of pre-LMXBs consisting a $1.4~M_{\odot}$ NS and a $1.5~M_{\odot}$ MS companion in the orbital periods vs. donor-star masses diagram. The solid, dashed, and dotted curves correspond to an initial orbital period of 2.95, 2.88, and 2.8 days, respectively. The crosses denote eight UCXBs with observed distances. } \label{fig:orbmass}
\end{figure}

The radius of a WD with a mass $M_{\rm wd}$ is \citep{rapp87,tout97}
\begin{equation}
R_{\rm wd}=0.0115R_{\odot}\sqrt{\left(\frac{M_{\rm ch}}{M_{\rm wd}}\right)^{2/3}-\left(\frac{M_{\rm wd}}{M_{\rm ch}}\right)^{2/3}}, \label{rwd}
\end{equation}
where $M_{\rm ch}=1.44~M_{\odot}$ is the Chandrasekhar mass limit. Considering $M_{\rm wd}\ll M_{\rm ch}$ in UCXBs, Eq. (\ref{rwd}) can approximately be written as
\begin{equation}
R_{\rm wd}\approx 0.0115R_{\odot}\left(\frac{M_{\rm ch}}{M_{\rm wd}}\right)^{1/3}. \label{rwd2}
\end{equation}
When $M_{\rm wd}\leq 0.8M_{\rm ns}$, the Roche-lobe radius of the WD is \citep{pacz71}
\begin{equation}
R_{\rm L}=0.462\left(\frac{M_{\rm wd}}{M_{\rm ns}+M_{\rm wd}}\right)^{1/3}a,\label{rwd3}
\end{equation}
where $a$ is the orbital separation. Because the WD companion would fill its Roche lobe during the UCXBs stage, $R_{\rm L}=R_{\rm wd}$. From Eqs. (\ref{rwd2}) and (\ref{rwd3}) combined with the Kepler's third law, it yields a WD mass-orbital period relation of UCXBs as
\begin{equation}
M_{\rm wd}\approx \frac{47.2~\rm s}{P_{\rm orb}}~M_{\odot}.\label{mwd}
\end{equation}
This relation is approximately consistent with that derived by \cite{rapp87} as $M_{\rm wd}=46~{\rm s}/P_{\rm orb}~M_{\odot}$ \citep{prod15}.

\subsection{Formation and detectability of eight UCXBs}

At present, there are about fifteen known UCXBs in the Galaxy, including ten persistent sources and five transient sources \citep{hein13}. To test the evolutionary channel of UCXBs, we list eight UCXBs with the observed distances in Table 2. The WD masses are derived by the WD mass-orbital period relation. Figure 5 displays the evolution of pre-LMXBs consisting a $1.4~M_{\odot}$ NS and a $1.5~M_{\odot}$ MS companion with initial orbital periods of 2.95, 2.88, and 2.8 days in the orbital period versus donor-star mass diagram. It requires an initial orbital period equaling to the bifurcation period for pre-LMXBs to evolve into four observed UCXBs with orbital periods less than 23 min. For four UCXBs with long orbital periods in the range 43 - 50 min, the initial orbital periods of pre-LMXBs should be in a fine-tuning range near the bifurcation period. Therefore, the standard magnetic braking scenario is successful in reproducing the observed UCXBs \citep{rapp83}.

\begin{figure}
\centering
\includegraphics[width=1.15\linewidth,trim={0 0 0 0},clip]{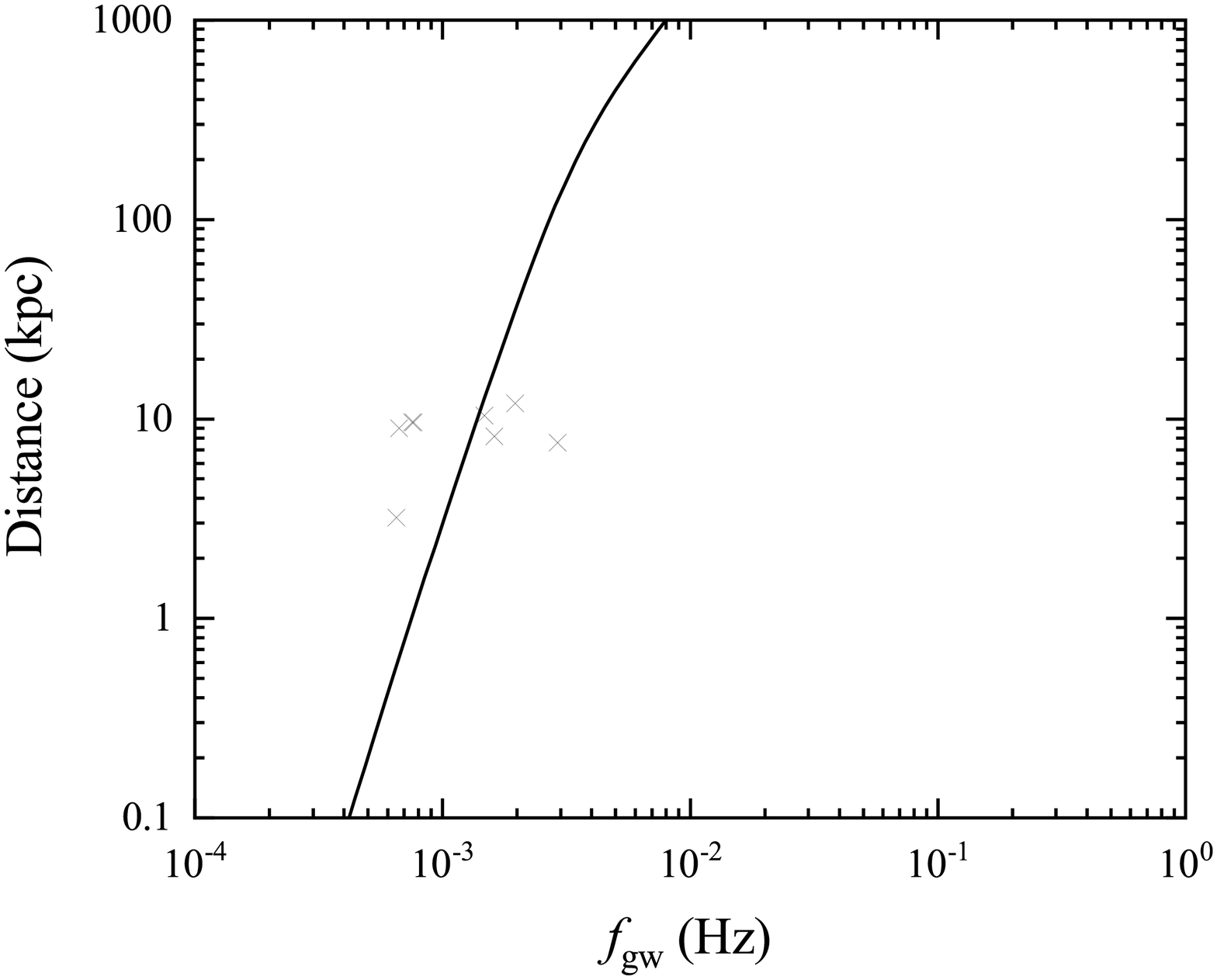}
\caption{Maximum distances of UCXBs detecting by the LISA as a function of GW frequency. The NS masses in UCXBs are assumed to be $1.6~M_{\odot}$. The crosses denote eight UCXBs with the observed distances. } \label{fig:orbmass}
\end{figure}

Considering the mass growth of the accreting NS during the recycled stage, the NS masses in UCXBs are assumed to be $1.6~M_{\odot}$. We then can estimate the chirp mass of UCXBs by
\begin{equation}
\mathcal{M}=\frac{(M_{\rm ns}M_{\rm d})^{3/5}}{(M_{\rm ns}+M_{\rm d})^{1/5}}.
\end{equation}
According to Equation (1), the characteristic strain of eight UCXBs can be derived. In Figure 3, we also plot the locations of eight UCXBs in the characteristic strain amplitude versus GW frequency diagram. Four sources with short orbital periods can be detected by the LISA, and the locations of seven sources are consistent with the evolutionary tracks when $d=10~\rm kpc$. However, the source 4U0614$+$091 is exceptive. Its optical spectrum confirmed the donor star to be a CO WD \citep{nele04a,nele06}, which cannot be produced by the evolutionary channel we studied.

\begin{table}
\begin{center}
\caption{The observed and derived parameters of eight UCXBs. \label{tbl-2}}
\begin{tabular}{@{}llllll@{}}
\hline\hline\noalign{\smallskip}
Source &$P_{\rm orb}$&$M_{\rm wd}$ & $d$ &location& Reference \\
 & (s) &  ($ M_{\odot}$) & (kpc)&  &\\
\hline\noalign{\smallskip}
4U1820$-$30 & 685  &0.069 &7.6& NGC6624 &1,2\\
4U0513$-$40  & 1020 &0.046 &12& NGC1851 &3,4\\
4U1850$-$087 & 1236 &0.038 &8.2&NGC6712  &5\\
M15 X$-$2    & 1356 &0.035 &10.4&M15 &3,6\\
NGC6652B     & 2616 &0.018 &9.6 &NGC6652 &  7,8  \\
XB1832$-$330  & 2628 &0.018 &9.6 &NGC6652  & 9   \\
4U1915$-$05  & 3000 &0.016 &9.0 &  &  10  \\
4U0614$+$091 & 3060 &0.015 &3.2 &  & 11,12   \\
\hline\noalign{\smallskip}
\end{tabular}
\tablenotetext{}{}
\\{References. (1) \cite{stel87}, (2) \cite{guve10}, (3) \cite{harr96}, (4) \cite{zure09}, (5) \cite{home96},
 (6) \cite{dieb05}, (7) \cite{hein01}, (8) \cite{int07}, (9) \cite{deut00}, (10) \cite{whit82}, (11) \cite{nele04a}, (12) \cite{shah08}. }
\\{Note. The WD masses are derived by the equation (6).}
\end{center}
\end{table}

\subsection{Maximum detectable distances of UCXBs by the LISA}
In theory, the relevant UCXBs will be detectable by the LISA if their characteristic strains are greater than the sensitive strains of the LISA. According to Equation (\ref{mwd}), the mass of the WD is $M_{\rm wd}=23.6{\rm s}~f_{\rm gw}~M_{\odot}$. Taking $M_{\rm ns}=1.6~M_{\odot}$, from $h_{\rm c}(f_{\rm gw})>h_{\rm min}(f_{\rm gw})$ (here $h_{\rm min}$ is the sensitive strains of the LISA), the maximum detectable distance of UCXBs by the LISA is given by
\begin{eqnarray}
d_{\rm max}=15{\rm kpc} \frac{2.5\times 10^{-20}}{h_{\rm min}(f_{\rm gw})}\left(\frac{f_{\rm gw}}{1~\rm mHz}\right)^{7/6}\nonumber\\
\frac{1.6\times23.6f_{\rm gw}}{(1.6+23.6f_{\rm gw})^{5/3}}.
\end{eqnarray}

In Figure 6, we plot the maximum detectable distances of UCXBs by the LISA for different GW frequencies. UCXBs below the curve
can be detected by the LISA, hence four sources with short orbital periods among eight known UCXBs are potential LISA sources. According to this curve, an UCXB with a GW frequency of 0.0014 Hz (corresponding to an orbital period of 24 min) is detectable by the LISA at a distance of 10 kpc, while the maximum detectable distance of GW signals of 0.008 Hz (corresponding to an orbital period of 4.1 min) can reach 1 Mpc.

\section{Binary Population synthesis}
By employing the rapid binary evolution code presented by \cite{Hur02}, we carried out a series of binary population synthesis (BPS) simulations to investigate the Galactic birthrate of UCXB-LISA sources.
A sample of $1\times10^{\rm 7}$ primordial binaries are evolved until the formation of NS$+$MS star systems in each simulation. We assume that a UCXB would be produced when the parameters of the NS$+$MS system locate in the parameter space of the pre-LMXBs/IMXBs evolving toward UCXB-LISA sources in Figure\,4.

The initial parameters and basic assumptions in the Monte Carlo BPS computations are shown as follows:

(1) For the primordial primary masses, we adopt the initial mass function presented by \cite{mil79}.

(2) A constant mass ratio distribution is employed, i.e., $n(q)=1$, in which $0<q\le1$.

(3) The initial distribution of separations $a$ is supposed to be constant in $\log a$ for wide binaries with orbital periods larger than 100 yr, and fall off smoothly for close binaries (see Eggleton et al. 1989).

(4) All stars are assumed to be members of binary systems with circular orbits.

(5) The star formation rate is adopted to be constant ($5\, M_{\odot}\rm yr^{\rm -1}$) for the Galaxy over the past $15\,\rm Gyr$ \citep{YL98,Will04,han04}.

(6) We adopt the standard energy prescription from \cite{webb84} to approximate the common envelope (CE) ejection process, and the uncertain parameters $\alpha_{\rm CE}$ and $\lambda$ in this prescription are combined as a single parameter and set to be $\alpha_{\rm CE}\lambda=0.5$ and 1.5 for a comparison.

\begin{figure}
\centering
\includegraphics[angle=0,width=1.15\linewidth,trim={0 0 0 0},clip]{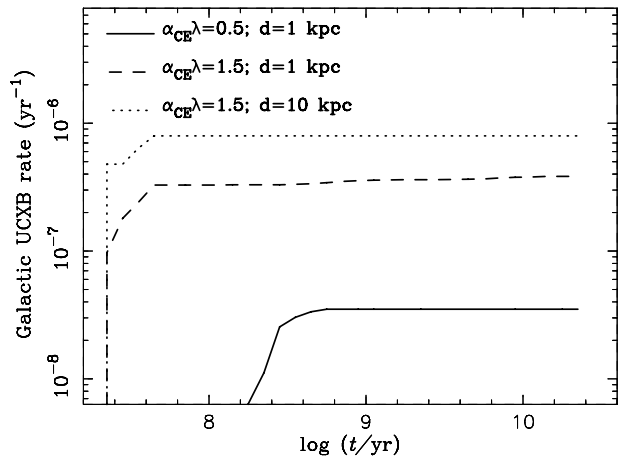}
\caption{Evolution of Galactic UCXB-LISA sources birthrates as a function of time by adopting a constant star formation rate of $5~\rm M_{\odot} yr^{\rm -1}$. The results of $d=1$ kpc, and 10 kpc are obtained by simply using corresponding parameter spaces shown in Figure 4 and normalizing to the corresponding volume based on Equation (9).} \label{fig:orbmass}
\end{figure}

Because NS+MS systems in the relevant parameter space will evolve into UCXB-LISA sources, our simulated birthrates also represent the birthrates $\mathcal{R}(t)$ of UCXB-LISA sources in the Galaxy. Taking the radius and the scale height of the Galaxy are 15 kpc and 1 kpc, and assuming that the binary systems obey a uniform distribution in the Galactic disk, the birthrate of UCXB-LISA sources within the specific distance $d$ is
\begin{equation}
\mathcal{R}_{\rm d}(t)=\mathcal{R}(t)\left(\frac{d}{15~\rm kpc}\right)^{2}.
\end{equation}

Figure 7 shows the birthrate evolution of pre-LMXBs/IMXBs that can evolve into UCXB-LISA sources in the Galaxy by adopting the parameter space of $d=1 ~\rm kpc$ and $d=10 ~\rm kpc$.  There exists a time delay between the NS+MS systems and the UCXBs ($\sim6-11~\rm Gyr$ , see also Table 1). For the case adopting the parameter space of $d=1 ~\rm kpc$, the Galactic birthrate of UCXBs are $(3-4)\times10^{\rm -7}$ and $3.5\times10^{\rm -8}\,\rm yr^{\rm -1}$ when $\alpha_{\rm CE}\lambda=1.5$, and 0.5, respectively. Because the CE ejection events with large $\alpha_{\rm CE}\lambda$ would be easy to happen for the release of the same orbital energy, producing more NS$+$MS systems located in the region of the pre-LMXBs/IMXBs in Figure 4. As a result, the birthrates of UCXBs increase with the value of the CE ejection parameter. For $d=10 ~\rm kpc$, the Galactic birthrate of UCXB-LISA sources is about $8.0\times10^{\rm -7}\,\rm yr^{\rm -1}$ when $\alpha_{\rm CE}\lambda=1.5$, while the birthrate is negligible when $\alpha_{\rm CE}\lambda=0.5$.

We then estimate the birthrate of UCXB-LISA sources in the Galaxy when $\alpha_{\rm CE}\lambda=1.5$. In the distance interval from 9 kpc to 10 kpc, the birthrate of UCXB-LISA sources is $8.0\times10^{\rm -7}(\frac{10^{2}-9^{2}}{10^{2}})~\rm yr^{\rm -1}\approx 1.6\times10^{\rm -7}~\rm yr^{-1}$. The birthrate of UCXB-LISA sources per kpc interval from i kpc to i+1 kpc is given by
\begin{equation}
\mathcal{R}_{\rm i,i+1}(t)=\mathcal{R}_{0,1}(t)P_{\rm i,i+1}[(i+1)^{2}-i^{2}],
\end{equation}
where $\mathcal{R}_{0,1}(t)=(3-4)\times10^{\rm -7}~\rm yr^{-1}$ is the birthrate of a distance range from 0 kpc to 1 kpc, $P_{\rm i,i+1}$ represents the ratio between the parameter space of the progenitors of UCXB-LISA sources for the specific kpc interval and that from 0 kpc to 1 kpc. In Figure 4, the mean orbital-period width of the progenitors of UCXB-LISA sources in a distance range from 9 kpc to 10 kpc is roughly 0.06 day, while the mean width from 0 kpc to 1 kpc is approximate 3 days. Assuming that initial companion masses and initial
orbital periods obey a uniform distribution, we then can approximately estimate $P_{\rm 9,10}\approx \frac{0.06}{3.0}=1/50$. According to Equation (10), $\mathcal{R}_{\rm 9,10}(t)\approx(3-4)\times10^{\rm -7}~\rm yr^{-1}\times20\times1/50=(1.2-1.6)\times10^{\rm -7}~\rm yr^{-1}$, which is in good agreement with the above calculation. For simplicity, we assume $P_{\rm i,i+1}\propto1/i$, and then Equation (10) yields a constant birthrate for $\mathcal{R}_{\rm i,i+1}(t)$ ($i> 0$). Therefore, the birthrate of UCXB-LISA sources in the Galaxy $\mathcal{R}_{0,15}(t)\approx\mathcal{R}_{0,1}(t)+14\mathcal{R}_{9,10}(t)=(2-2.6)\times10^{\rm -6}~\rm yr^{-1}$.

Note that there are some other BPS studies on the formation of UCXBs (e.g. Belczynski \& Taam 2004, Zhu et. al 2012, van haaften et al. 2013). Similar to the present work, the prescription for the stellar evolution in these studies also adopted the analytic formulas given by Hurley et al. (2000). However, these BPS studies directly evolve primordial binaries to the accreting BHs/NSs in close binaries with orbital periods shorter than 80 min (Belczynski \& Taam 2004) or 1 hr (van haaften et al. 2013), or to the accreting NS+WD/He star systems with orbital periods shorter than 1 hour (Zhu et al. 2012). In this work, we firstly obtained the parameter space of NS+MS systems that can evolve into UCXB-LISA sources, and then evolve primordial binaries to NS+MS systems that are located in the parameter space based on a BPS approach. Moreover, these other works only studied the properties of all UCXBs, while this work mainly focused on the investigation of the detectability of UCXBs by the LISA.

\section{Discussion and Conclusion}
UCXBs are generally thought to be promising GW sources that will be detected by the LISA detector \citep{nele09}, hence the detectability of this population is important for the future LISA mission. This work aims at the formation and birthrate of UCXB-LISA sources evolving from LMXBs/IMXBs in the Galaxy. Employing the MESA code, we simulate the evolution of a great number of pre-LMXBs/IMXBs consisting of a NS and a MS companion, and find that the systems with initial orbital periods lower than the bifurcation periods can evolve into UCXBs under an assumption of the standard magnetic braking. When $M_{\rm ns}=1.4~M_{\odot}$, $M_{\rm d,i}=0.4-3.3~M_{\odot}$, most pre-LMXBs/IMXBs with initial orbital periods lower than the bifurcation periods can evolve into UCXBs that are visible as LISA sources within a distance of $d=1~\rm kpc$. However, the pre-LMXBs/IMXBs require a fine-tuning initial orbital periods in order to form UCXBs appearing as LISA sources within 10 kpc. Furthermore, the initial orbital periods of pre-LMXBs/IMXBs almost equal to the bifurcation periods in order to evolve into detached pre-UCXBs consisting of a radio MSP and a He WD.

In the Galaxy, there exist eight known UCXBs with the observed distances. According to the WD mass-orbital period relation, the donor-star masses are derived. To evolve toward the current stage, the initial orbital periods of four UCXBs with short orbital periods should equal to the bifurcation period, while another four sources require initial orbital periods slightly lower than the bifurcation period. Based on the WD mass-orbital period relation, we obtain the maximum detectable distances of UCXB-LISA sources for specific GW frequencies. The observed distances of four sources in eight UCXBs are within the maximum detectable distance, hence they are potential LISA sources. In the theory, the maximum detectable distance of an UCXB with an orbital period of 4.1 min can reach 1 Mpc. However, it is difficult to form such an UCXB in our simulations.

Based on the initial parameter space of pre-LMXBs/IMXBs that can evolve into UCXB-LISA sources, we calculate their birthrates by the rapid binary evolution code developed by \cite{Hur02}. When the CE parameter $\alpha_{\rm CE}\lambda=1.5$, the birthrate of UCXB-LISA sources are $(3-4)\times10^{\rm -7}$ and $8\times10^{-7}~\rm yr^{\rm -1}$ within a distance of 1 kpc, and 10 kpc, respectively. Adopting a simply assumption that the initial parameter space is inversely proportional to the distance, the birthrate of UCXB-LISA sources in the Galaxy can be estimated to be $\mathcal{R}_{0,15}(t)\approx(2-2.6)\times10^{\rm -6}~\rm yr^{-1}$. If we take the timescale of UCXBs appearing as LISA sources to be 30 Myr (see also Table 1), there exist about 60 - 80 UCXB-LISA sources in the Galaxy. This estimation is slightly smaller than the derived minimum number (100) based on the known numbers of binary radio MSPs \citep{taur18}, in which the LISA sources include detached pre-UCXBs consisting a radio MSP and a He WD.

Our population synthesis simulations focus on the formation and evolution of UCXBs from LMXBs/IMXBs in the Galactic field. Therefore, our results only represent a lower limit of UCXB-LISA sources in the Galaxy. However, owing to a dense stellar density, it is easy to form UCXBs in the globular clusters by the dynamic processes including direct collisions \citep{verb87,rasi91,davi92,ivan05,lomb06}, tidal captures \citep{bail87,pods02,voss07}, and exchange interactions \citep{davi98,rasi00,ivan10}. Six sources among eight known UCXBs with the observed distances actually locate in the globular clusters. By assuming that the ratio between the number of UCXBs in the Galactic field and those in the globular clusters is $1/4$, the number of UCXB-LISA sources in the Galaxy can be estimated to be 240 - 320. This number is significantly smaller than the predicted number of AM CVn stars \citep[$\sim 10^{4},$][]{nele04b} and detached double WD systems \citep[$\sim 10^{4}$,][]{nele01a,liu10,ruit10,yu10}, and obviously larger than that of intermediate-mass black hole X-ray binaries \citep[$\la10$,][]{chen20} that can be detectable by the LISA in the Galaxy. However, the UCXB-LISA sources provide an opportunity to pursue full multi-messenger investigations. Therefore, the detection of UCXBs as LISA sources is still significant.

The birthrate and number of UCXB-LISA sources mentioned above are based on the assumption that the metallicity of the donor stars is $Z=0.02$. If the donor stars possess a low metallicity, our simulations indicate that the initial parameter space of the progenitors of UCXB-LISA sources would obviously shrink. When the initial mass of the donor star $M_{\rm d,i}=1.5~M_{\odot}$, the initial orbital periods of pre-LMXBs that can evolve into UCXB-LISA sources within 1 kpc are in the range of 0.6 - 2.4 days, and 0.6 - 1.4 days for $Z=0.01$, and 0.001, respectively. The shrinkage of the parameter space would result in a decrease of the birthrate \cite[e.g.][]{wang10}. Therefore, a low metallicity of the donor stars tends to produce small number and low birthrate of UCXB-LISA sources.

\acknowledgments {We thank the referee for a very careful reading and constructive comments that have led to the improvement of the manuscript. WCC was partly supported by the National Natural Science Foundation of China (under grant Nos. 11573016, and 11733009), the Program for Innovative Research Team (in Science and Technology) at the
University of Henan Province. DDL was supported by the National Natural Science Foundation of China (No. 11903075) and the Western Light Youth Project of Chinese Academy of Sciences. BW was supported by the National Natural Science Foundation of China (Nos 11873085, 11673059 and 11521303), the Chinese Academy of Sciences (No QYZDB-SSW-SYS001),
and the Yunnan Province (Nos 2018FB005 and 2019FJ001).}

If you require the inlists for the MESA simulations in this work, please contact chenwc@pku.edu.cn.


\begin{thebibliography}{}
\bibitem[Abbott et al. (2016)]{abbo16} Abbott, B. P., Abbott, R., Abbott, T. D., et al. 2016, PhRvL, 116, 061102
\bibitem[Abbott et al. (2017)]{abbo17} Abbott, B. P., Abbott, R., Abbott, T. D., et al. 2017, PhRvL, 119, 161101
\bibitem[Amaro-Seoane et al. (2017)]{amar17} Amaro-Seoane, P.,  Audley, H., Babak, S., et al., 2017, arXiv:1702.00786
\bibitem[Bailyn \& Grindlay (1987)]{bail87} Bailyn, C. D., \& Grindlay, J. E. 1987, ApJL, 316, L25
\bibitem[Belczynski \& Taam (2004)]{belc04} Belczynski, K., \& Taam, R. E. 2004, ApJ, 603, 690
\bibitem[Cartwright et al. (2013)]{cart13} Cartwright, T. F., Engel, M. C., Heinke, C. O., et al. 2013, ApJ, 768, 183
\bibitem[Chen \& Podsiadlowski (2016)]{chen16} Chen, W.-C., \& Podsiadlowski, P. 2016, ApJ, 830, 131
\bibitem[Chen (2020)]{chen20} Chen, W.-C. 2020, ApJ, 896, 129
\bibitem[Davies et al. (1992)]{davi92} Davies, M. B., Benz, W., \& Hills, J. G. 1992, ApJ, 401, 246
\bibitem[Davies \& Hansen (1998)]{davi98} Davies, M. B., \& Hansen, B. M. S. 1998, MNRAS, 301, 15
\bibitem[Deutsch et al. (2000)]{deut00} Deutsch, E. W., Margon, B., \& Anderson, S. F. 2000, ApJL, 530, L21
\bibitem[Dieball et al. (2005)]{dieb05} Dieball, A., Knigge, C., Zurek, D. R., et al. 2005, ApJL, 634, L105
\bibitem[Eggleton, Fitchett \& Tout (1989)]{Egg89} Eggleton, P. P., Fitchett, M. J., \& Tout, C. A. 1989, ApJ, 347, 998
%\bibitem[Ergma et al. (1998)]{ergm98} Ergma, E., Sarna, M. J., \& Antipova, J. 1998, MNRAS, 300, 352
\bibitem[Ergma (1996)]{ergm96a} Ergma, E. 1996, A\&A, 315, L17
\bibitem[Ergma \& Sarna (1996)]{ergm96b} Ergma, E., \& Sarna, M. J. 1996, MNRAS, 280, 1000
\bibitem[G\"{u}ver et al. (2010)]{guve10} G\"{u}ver, T., Wroblewski, P., Camarota, L., \& \"{O}zel, F. 2010, ApJ, 719, 1807
\bibitem[Han \& Podsiadlowski (2004)]{han04} Han, Z., \& Podsiadlowski, Ph. 2004, MNRAS, 350, 1301
\bibitem[Harris (1996)]{harr96} Harris, W. E. 1996, AJ, 112, 1487
\bibitem[Heinke et al. (2001)]{hein01} Heinke, C. O., Edmonds, P.D., \& Grindlay, J. E. 2001. ApJ 562, 363
\bibitem[Heinke et al. (2013)]{hein13} Heinke, C. O., Ivanova, N., Engel, M. C., et al. 2013, ApJ, 768, 184
\bibitem[Hils \& Bender (2000)]{hils00} Hils, D., \&  Bender, P. L. 2000, ApJ, 537, 334
\bibitem[Hurley et al. (2002)]{Hur02} Hurley, J. R., Tout, C. A., \& Pols, O. R. 2002, MNRAS, 329, 897
\bibitem[Homer et al. (1996)]{home96} Homer, L., Charles, P. A., Naylor, T., et al. 1996, MNRAS, 282, L37
\bibitem[Iben et al. (1995)]{iben95} Iben, I. J., Tutukov, A. V., \& Yungelson, L. R. 1995, ApJS, 100, 233
\bibitem[in't Zand et al. (2007)]{int07} in't Zand, J. J. M., Jonker, P. G., \& Markwardt, C. B. 2007, A\&A, 465, 953
\bibitem[Istrate et al. (2014a)]{istr14a} Istrate, A. G., Tauris, T. M., Langer, N., \& Antoniadis, J., 2014a, A\&A, 571, L3
\bibitem[Istrate et al. (2014b)]{istr14b} Istrate, A. G., Tauris, T. M., \& Langer, N. 2014b, A\&A, 571, A45

\bibitem[Ivanova et al. (2010)]{ivan10} Ivanova, N., Chaichenets, S., Fregeau, J., et al. 2010, ApJ, 717, 948
\bibitem[Ivanova et al. (2013)]{ivan13} Ivanova, N., Justham, S., Chen, X., et al. 2013, A\&ARv, 21, 59
\bibitem[Ivanova et al. (2005)]{ivan05} Ivanova, N., Rasio, F. A., Lombardi, J. C., Jr., Dooley, K. L., \& Proulx, Z. F.
2005, ApJL, 621, L109
\bibitem[Kaplan et al. (2012)]{kapl12} Kaplan, D. L., Bildsten, L. \&  Steinfadt, J. D. R. 2012, ApJ, 758, 64
\bibitem[Kremer et al. (2017)]{krem17} Kremer, K., Breivik, K., Larson, S. L., \& Kalogera, V. 2017, ApJ, 846, 95

\bibitem[Lin et al. (2011)]{lin11} Lin, J., Rappaport, S., Podsiadlowski, P., et al. 2011, ApJ, 732, 70
\bibitem[Liu \& Wang (2020)]{liu20} Liu, D., \& Wang, B. 2020, MNRAS, 494, 3422
\bibitem[Liu et al. (2010)]{liu10} Liu, J., Han, Z., Zhang, F., \& Zhang, Y. 2010, ApJ, 719, 1546
\bibitem[L\"{u} et al. (2017)]{lu17} L\"{u}, G., Zhu, C., Wang, Z., \& Iminniyaz, H. 2017, ApJ, 847, 62
\bibitem[Lombardi et al. (2006)]{lomb06} Lombardi, J. C., Jr, Proulx, Z. F., Dooley, K. L., et al. 2006, ApJ, 640, 441
\bibitem[Ma \& Li (2009a)]{ma09a} Ma, B., \& Li, X.-D. 2009a, ApJ, 691, 1611
\bibitem[Ma \& Li (2009b)]{ma09b} Ma, B., \& Li, X.-D. 2009b, ApJ, 698, 1907
\bibitem[Miller \& Scalo (1979)]{mil79} Miller, G. E., \& Scalo, J. M. 1979, ApJS, 41, 513
\bibitem[Nelemans (2003)]{nele03} Nelemans, G. 2003, CQGra, 10, 81
\bibitem[Nelemans (2009)]{nele09} Nelemans, G. 2009, CQGra, 26, 094030
\bibitem[Nelemans (2013)]{nele13} Nelemans, G. 2013, in Auger, G., Bin\'{e}truy, P., Plagnol, E., eds, ASP Conf. Ser. Vol. 467, 9th LISA Symposium. Astron. Soc. Pac., San Francisco, 27
\bibitem[Nelemans \& Jonker (2010)]{nele10} Nelemans, G., \& Jonker, P. G. 2010, New Astronomy Review, 54, 87
\bibitem[Nelemans et al. (2004a)]{nele04a} Nelemans, G., Jonker, P. G., Marsh, T. R., \& van der Klis, M. 2004a, MNRAS, 348, L7
\bibitem[Nelemans et al. (2006)]{nele06} Nelemans, G., Jonker, P. G., \& Steeghs, D. 2006, MNRAS, 370, 255
\bibitem[Nelemans et al. (2001b)]{nele01b} Nelemans, G., Portegies Zwart, S. F., Verbunt, F., \& Yungelson, L. R. 2001b, A\&A, 368, 939
\bibitem[Nelemans et al. (2000)]{nele00} Nelemans, G., Verbunt, F., Yungelson, L. R. , \& Portegies Zwart, S. F. 2000, A\&A, 360, 1011
\bibitem[Nelemans et al. (2001a)]{nele01a} Nelemans, G., Yungelson, L. R. , \& Portegies Zwart, S. F. 2001a, A\&A, 375, 890
\bibitem[Nelemans et al. (2001c)]{nele01c} Nelemans, G., Yungelson, L. R., Portegies Zwart, S. F., \& Verbunt, F. 2001c, A\&A, 365, 491
\bibitem[Nelemans et al. (2004b)]{nele04b} Nelemans, G., Yungelson, L. R., \& Portegies Zwart, S. F. 2004b, MNRAS, 349, 181


\bibitem[Nelson (1986)]{nels86} Nelson, L. A., Rappaport, S. A., \& Joss, P. C., 1986, ApJ, 304, 231
\bibitem[Paczy\'{n}ski (1971)]{pacz71} Paczy\'{n}ski, B. 1971, ARA\&A, 9, 183
\bibitem[Paxton et al. (2015)]{paxt15} Paxton, B., Marchant, P., Schwab, J., et al. 2015, ApJS, 220, 15
\bibitem[Podsiadlowski et al. (2002)]{pods02} Podsiadlowski, P., Rappaport, S., \& Pfahl, E. D. 2002, ApJ, 565, 1107
\bibitem[Podsiadlowski et al. (2003)]{pods03} Podsiadlowski, P., Rappaport, S., \& Han, Z. 2003, MNRAS, 341, 385
\bibitem[Prodan \& Murray (2015)]{prod15} Prodan, S., \& Murray, N. 2015, ApJ, 798, 117
\bibitem[Pylyser \& Savonije (1988)]{pyly88} Pylyser, E. H. P., \& Savonije, G. J. 1988, A\&A, 191, 57
%\bibitem[Pylyser \& Savonije (1989)]{pyly89} Pylyser, E. H. P., \& Savonije, G. J. 1989, A\&A, 208, 52


\bibitem[Rappaport et al. (1983)]{rapp83} Rappaport, S., Verbunt, F., \& Joss, P. C. 1983, ApJ, 275, 713
\bibitem[Rappaport et al. (1987)]{rapp87} Rappaport, S., Ma, C. P., Joss, P. C., \& Nelson, L. A. 1987, ApJ, 322, 842
\bibitem[Rasio \& Shapiro (1991)]{rasi91} Rasio, F. A., \& Shapiro, S. L. 1991, ApJ, 377, 559
\bibitem[Rasio et al. (2000)]{rasi00} Rasio, F. A., Pfahl, E. D., \& Rappaport, S. 2000, ApJL, 532, L47
\bibitem[Ruiter et al. (2010)]{ruit10} Ruiter, A. J., Belczynski, K., Benacquista, M., Larson, S. L., \& Williams, G. 2010, ApJ, 717, 1006
\bibitem[Robson et al. (2018)]{robs18} Robson, T., Cornish, N. J., \& Liu, C. 2018, arXiv: 1803.01944
\bibitem[Sengar et al. (2017)]{seng17} Sengar, R., Tauris, T. M., Langer, N., \& Istrate, A. G. 2017, MNRAS, 470, L6
\bibitem[Shahbaz et al. (2008)]{shah08} Shahbaz, T., Watson, C. A., Zurita, C., et al. 2008, PASP, 120, 848
\bibitem[Shao \& Li (2012)]{shao12} Shao, Y., \& Li, X. -D. 2012, ApJ, 756, 85
\bibitem[Stella et al. (1987)]{stel87} Stella, L., Priedhorsky, W., \& White, N. E. 1987, ApJL, 312, L17
\bibitem[Taam \& Sandquist (2000)]{taam00} Taam, R. E., \& Sandquist, E. L. 2000, ARA\&A, 38, 113
\bibitem[Tauris \& van den Heuvel (2006)]{taur06} Tauris, T. M., \& van den Heuvel, E. P. J. 2006, in Formation and Evolution
of Compact Stellar X-ray Sources, ed. W. H. G. Lewin, \& M. van der Klis
(Cambridge: Cambridge Univ. Press), 623
\bibitem[Tauris (2018)]{taur18} Tauris, T. M. 2018, PhRvL, 121, 131105
\bibitem[Tauris et al. (2000)]{taur00} Tauris, T. M., van den Heuvel, E. P. J., \& Savonije, G. J. 2000, ApJL, 530, L93
\bibitem[Tauris et al. (2017)]{taur17} Tauris, T. M.,  Kramer,  M.,  Freire, P. C. C., et al. 2017, ApJ, 846, 170
\bibitem[Tout et al. (1997)]{tout97} Tout, C. A., Aarseth, S. J., Pols, O. R., et al. 1997, MNRAS, 291, 732
\bibitem[Tutukov \& Yungelson (1979)]{tutu79} Tutukov, A. V., \& Yungelson, L. R. 1979, AcA, 29, 665
\bibitem[Tutukov et al. (1985)]{tutu85} Tutukov, A. V., Fedorova, A. V., Ergma, E. V., \& Yungelson, L. R. 1985, SvAL, 11, 52
\bibitem[Tutukov et al. (1987)]{tutu87} Tutukov, A. V., Fedorova, A. V., Ergma, E. V., \& Yungelson, L. R. 1987, SvAL, 13, 328
\bibitem[Tutukov \& Yungelson (1993)]{tutu93} Tutukov, A. V., \& Yungelson, L. R. 1993, ARep, 37, 411
\bibitem[van der Sluys et al. (2005a)]{sluy05a} van der Sluys, M. V., Verbunt, F., \& Pols, O. R. 2005a, A\&A, 431, 647
\bibitem[van der Sluys et al. (2005b)]{sluy05b} van der Sluys, M. V., Verbunt, F., \& Pols, O. R. 2005b, A\&A, 440, 973
\bibitem[van der Sluys et al. (2006)]{sluy06} van der Sluys, M. V., Verbunt, F., \& Pols, O. R. 2006, A\&A, 460, 209
\bibitem[van der Sluys (2011)]{sluy11} van der Sluys, M. 2011, in Evolution of Compact Binaries, edited by L. Schmidtobreick, M. R. Schreiber, \& C. Tappert, vol. 447 of Astronomical Society of the Pacific Conference Series, 317
\bibitem[van Haaften et al. (2012)]{haaf12} van Haaften, L. M., Nelemans, G., Voss, R., Wood, M. A., \& Kuijpers, J.
2012, A\&A, 537, A104
\bibitem[van Haaften et al. (2013)]{haaf13} van Haaften, L. M., Nelemans, G., Voss, R., Toonen, S., Portegies Zwart, S. F., Yungelson, L. R., \& van der Sluys, M. V. 2013, A\&A, 552, 69
\bibitem[Verbunt (1987)]{verb87} Verbunt, F. 1987, ApJL, 312, L23
\bibitem[Voss \& Gilfanov (2007)]{voss07} Voss, R., \& Gilfanov, M. 2007, MNRAS, 380, 1685
\bibitem[Wang \& Han (2010)]{wang10} Wang, B., \& Han, Z. 2010, A\&A, 515, A88
\bibitem[Webbink (1984)]{webb84} Webbink, R. F. 1984, ApJ, 277, 355
\bibitem[White \& Swank (1982)]{whit82} White, N. E., \& Swank, J. H. 1982, ApJ, 253, L61
\bibitem[Willems \& Kolb (2004)]{Will04} Willems, B., \& Kolb, U. 2004, A\&A, 419, 1057
\bibitem[Yungelson \& Livio (1998)]{YL98} Yungelson, L. R., \& Livio, M. 1998, ApJ, 497, 168
\bibitem[Yungelson et al. (2002)]{yung02} Yungelson, L. R., Nelemans, G., \& van den Heuvel, E. P. J. 2002, A\&A, 388, 546
\bibitem[Zhu et al. (2012)]{zhu12} Zhu, C.-H., L\"{u}, G.-L., \& Wang, Z.-J. 2012, RAA, 12, 1526
\bibitem[Zurek et al. (2009)]{zure09} Zurek, D. R.,Knigge, C., Maccarone, T. J., Dieball, A., \& Long, K. S. 2009, ApJ,
699, 1113
\bibitem[Yu \& Jeffery (2010)]{yu10} Yu, S., \& Jeffery, C. S. 2010, A\&A, 521, A85
\bibitem[Yu \& Jeffery (2015)]{yu15} Yu, S., \& Jeffery, C. S. 2015, MNRAS, 448, 1078

\end{thebibliography}
\end{document}